\documentclass{LMCS}

\usepackage{amssymb}
\usepackage{latexsym}
\usepackage{amsmath}
\usepackage{amsopn}
\usepackage{color}
\usepackage{pstricks}
\usepackage{pst-node}
\usepackage{pst-tree}
\usepackage{hyperref}

\newcommand{\nC}{\newcommand}

\nC{\theo}{\begin{thm}}
\nC{\theorem}{\begin{thm}}
\nC{\etheo}{\end{thm}}

\nC{\lemma}{\begin{lem}}
\nC{\elem}{\end{lem}}
\nC{\elemma}{\end{lem}}

\nC{\corol}{\begin{cor}}
\nC{\ecorol}{\end{cor}}

\nC{\defn}{\begin{defi}}
\nC{\edefn}{\end{defi}}

\nC{\define}[1]{{\bf\boldmath #1}}

\nC{\proposition}{\begin{prop}}
\nC{\eprop}{\end{prop}}

\newcounter{enum}
\newenvironment{enum}{\begin{list}{(\arabic{enum})}%
{\setlength{\labelwidth}{5mm}\setlength{\leftmargin}{10mm}%
\setlength{\itemindent}{0pt}\usecounter{enum}}}{\end{list}}

\newcounter{menum}

\newcounter{aenum}

\newcommand{\look}{\proof}

\newcommand{\hx}{\qed}

\newcommand{\usec}[1]{\par\medskip\noindent{\bf #1}}

\newcommand{\lra}{\longrightarrow}

\newcommand{\Iff}{\Longleftrightarrow}
\newcommand{\ra}{\rightarrow}

\renewcommand{\phi}{\varphi}
\renewcommand{\theta}{\vartheta}

\renewcommand{\epsilon}{\varepsilon}

\newcommand{\tle}{\lhd}

\newcommand{\N}{{\mathbb N}}

\nC{\IF}{{\bf if }}
\nC{\THEN}{{\bf then } }
\nC{\ELSE}{{\bf else } }
\nC{\DO}{{\bf do }}
\nC{\OD}{{\bf od } }
\nC{\OUTPUT}{{\bf output }}
\nC{\END}{{\bf end }}
\nC{\WHILE}{{\bf while} }
\nC{\INPUT}{{\bf Input} }
\nC{\GUESS}{{\bf guess } }
\nC{\CHOOSE}{{\bf universally choose } }
\nC{\REJECT}{{\bf reject } }
\nC{\ACCEPT}{{\bf accept } }
\nC{\FOR}{{\bf for } }
\nC{\TO}{{\bf to } }

\newcommand{\Ff}{{\mathcal F}}
\newcommand{\Gg}{{\mathcal G}}

\newcommand{\Kk}{{\mathcal K}}

\newcommand{\Pp}{{\mathcal P}}

\newcommand{\Tt}{{\mathcal T}}

\nC{\Exptime}{\text{\sc Exptime}}
\nC{\Nexptime}{\text{\sc Nexptime}}
\nC{\Ntime}{\text{\sc Ntime}}

\newcommand{\sig}{{\rm sig}}
\newcommand{\cutout}[1]{}

\newcommand{\incl}{\subseteq}
\newcommand{\mcnode}[2]{\circlenode{#1}{\strut #2}}
\newcommand{\mbnode}[2]{\rnode{#1}{\psframebox{\strut #2}}}
\newcommand{\set}[1]{\{#1\}}
\newcommand{\es}{\emptyset}

\newcommand{\wh}[1]{\widehat{#1}}
\newcommand{\wt}{\widetilde}

\newcommand{\w}{\omega}

\def\doi{2 (4:6) 2006}
\lmcsheading%
{\doi}
{1--22}
{}
{}
{Feb.~27, 2006}
{Nov.~\phantom{0}3, 2006}
{}

\begin{document}

\psset{nodesep=3pt,arrows=->}

\title[Positional Determinacy]{Positional Determinacy of Games with Infinitely 
Many Priorities\rsuper *}

\author[E.~Gr\"adel]{Erich Gr\"adel\rsuper a}
\address{{\lsuper a}Mathematische Grundlagen der Informatik, RWTH Aachen University, D-52056 Aachen, Germany}
\email{graedel@informatik.rwth-aachen.de}

\author[I.~Walukiewicz]{Igor Walukiewicz\rsuper b}
\address{{\lsuper b}LaBRI , Universit\'e  Bordeaux-1, 351  Cours de la Lib\'eration,
33 405 Talence, France}
\email{igw@labri.fr}

\keywords{Games, logic, positional determinacy, parity games, Muller games}
\subjclass{F.4.1, G2}
\titlecomment{{\lsuper *}This research has been partially supported by the
European Research Training Network ``Games and Automata for Synthesis 
and Validation'' (GAMES)}

\begin{abstract}
We study two-player games of infinite duration that are
played on finite or infinite game graphs. A winning strategy
for such a game is \emph{positional} if it only 
depends on the current position, and not
on the history of the play. 
A game is \emph{positionally determined}
if, from each position, one of the two players
has a positional winning strategy.

The theory of such games is well studied for winning conditions 
that are defined in terms of a mapping that assigns
to each position a \emph{priority} from a finite set $C$. 
Specifically, in Muller games the winner of a
play is determined by the set of those priorities
that have been seen infinitely often; an 
important special case are parity games where
the least (or greatest) priority 
occurring infinitely often determines the winner.   
It is well-known that parity games are
positionally determined whereas Muller games are determined via
finite-memory strategies.

In this paper, we extend this theory to the case
of games with infinitely many priorities.
Such games arise in several application areas,
for instance in pushdown games with winning conditions 
depending on stack contents.

For parity games there are several generalisations to the case of
infinitely many priorities.  While max-parity games over $\omega$ or
min-parity games over larger ordinals than $\omega$ require strategies
with infinite memory, we can prove that min-parity games
with priorities in $\omega$ are positionally determined.  Indeed, it
turns out that the min-parity condition over $\omega$ is
the only infinitary Muller condition that guarantees positional
determinacy on all game graphs.

\cutout{The discrepancy between (min-)parity games and max-parity
games has an intersting application to an old problem
posed by Jonathan Swift.}
\end{abstract}

\maketitle

\section{Motivation}\label{sec:motivation}

The problem of computing winning positions and
winning strategies in infinite games has numerous
applications in computing, most notably for the synthesis
and verification of reactive controllers  
and for the model-checking of the $\mu$-calculus and other logics.
Of special importance are \emph{parity games},
due to several reasons.
\begin{enum}
\item Many classes of games arising in practical applications admit
  reductions to parity games (over larger game graphs). This is the
  case for games modeling reactive systems, with winning conditions
  specified in some temporal logic or in monadic second-order logic
  over infinite paths (S1S), for Muller games, but also for games with
  partial information appearing in the synthesis of distributed
  controllers \cite{ArnoldVinWal03}.
\item Parity games arise as the model checking games for fixed point
  logics such as the modal $\mu$-calculus or LFP, the extension of
  first-order logic by least and greatest fixed points
  \cite{EmersonJutSis01,Graedel06}.  In particular the model checking
  problem for the modal $\mu$-calculus can be solved in polynomial
  time if, and only if, winning regions for parity games can be
  computed in polynomial time.
\item Parity games are positionally determined
  \cite{EmersonJut91,Mostowski91}. This means that from every
  position, one of the two players has a winning strategy whose moves
  depend only on the current position, not on the history of the
  play. This property is fundamental for numerous results in automata
  theory on infinite objects and for verification algorithms.
\end{enum} 

In most of the traditional applications of games in computer science,
the arena, and therefore also the number of priorities appearing in
the winning condition, are finite.  However, due to applications in
the verification of infinite-state systems and other areas where
infinite structures become increasingly important, it is interesting
to study infinite arenas that admit some kind of
finite presentation.  The best studied class of such games are
\emph{pushdown games} \cite{KupfermanVar00,Walukiewicz01}, where the
arena is the configuration graph of a pushdown automaton. Other
relevant classes of infinite, but finitely presented, (game) graphs
include prefix-recognizable graphs, HR- and
VR-equational graphs, graphs in the Caucal hierarchy, and automatic
graphs.  On all these classes of graphs (with the exception of
automatic graphs \cite{BlumensathGra04}), monadic second-order logic
can be evaluated effectively, which implies, for instance, that
winning regions of parity games with a finite number of priorities are
decidable.  However, once we move to infinite game graphs, winning
conditions depending on infinitely many priorities arise naturally. In
pushdown games, stack height and stack contents are natural parameters
that may take infinitely many values.  In \cite{CachatDupTho02},
Cachat, Duparc, and Thomas study pushdown games with an infinity
condition on stack contents, and Bouquet, Serre, and Walukiewicz
\cite{BouquetSerWal03} consider more general winning conditions for
pushdown games, combining a parity condition on the states of the
underlying pushdown automaton with an unboudedness condition on stack
heights. Similarly, Gimbert \cite{Gimbert04} considers games of
bounded degree where the parity winning conditions is combined with
the requirement that an infinite portion of the game graph is visited.

To establish positional determinacy or finite-memory determinacy
is a fundamental first step in the analysis of an infinite
game, and is also crucial for the algorithmic construction of winning
strategies. In the case of parity games 
with finitely many priorities the positional determinacy
immediately implies that winning regions can be
decided in NP $\cap$ Co-NP; with a little more effort it follows
that the problem is in fact in UP $\cap$ Co-UP 
\cite{Jurdzinski98}. 
Further, although it is not known yet whether parity games
can be solved in polynomial time, all known  
approaches towards an efficient algorithmic solution
make use of positional determinacy, including the presently
best deterministic algorithm from \cite{JurdzinskiPatZwi06}. 
The same is true for
the polynomial-time algorithms that we have for specific classes of 
parity games, including parity games with a bounded number of
priorities \cite{Jurdzinski00}, games where 
even and odd cycles do not intersect,
solitaire games and nested solitaire games 
\cite{BerwangerGra04}, and parity games of bounded tree width
\cite{Obdrzalek03}, bounded entanglement \cite{BerwangerGra05}, 
or bounded DAG-width \cite{BerwangerDawHunKre06,Obdrzalek06}.
Positional determinacy is also the key point in the proofs
of most of the known results on pushdown games.

In general, the positional determinacy of a game may depend 
on specific properties of the arena and on 
the winning condition. For instance, the 
previously known results
on pushdown games make use of the fact that
the arena is a pushdown graph.
However, this is not always the case. As we show here,
there are interesting cases, where positional determinacy 
is a consequence of the winning condition only.
Most notably this is the case for the parity condition
(little endian style) on $\omega$. 
In fact, we completely classify the 
infinitary Muller conditions with this property and 
show that they are equivalent to a parity condition.
This result gives a general, arena-independent 
explanation of the positional determinacy of certain pushdown games.
We hope and expect that it will be the first step for
algorithmic solutions for other infinite games with
finitely presented arenas.

\section{Introduction}

\subsection{Games and strategies}

We study two-player games of infinite duration
on arenas with infinitely many priorities.
An \emph{arena} $\Gg = ( V, V_0, V_1, E, \Omega )$,
consists of a directed graph $(V,E)$,
with a partioning $V=V_0\cup V_1$
of the nodes into positions of Player~0
and positions of Player~1.
The possible moves are described by
the edge relation
$E \subseteq V \times V$. The function
$\Omega: V\ra C$
assigns to every position a \emph{priority}.
Occasionally we encode the
priority function by the collection $(P_c)_{c\in C}$
of unary predicates where $P_c=\{v\in V: \Omega(v)=c\}$.

In case $(v,w)\in E$ we call $w$ a successor of $v$
and we denote the set of all successors of $v$ by $vE$.
To avoid tedious case distinctions,
we assume that every position has at least one successor.
A {\em play} of $\Gg$ is an infinite path $v_0v_1\dots $ formed by the two
players starting from a given initial position $v_0$.
Whenever the current position $v_n$ belongs to $V_\sigma$,
then Player~$\sigma$ chooses a successor $v_{n+1}\in v_n E$.
A \emph{game} is given by an arena and a \emph{winning condition}
that describes which of the  plays 
$v_0v_1\dots$ are won by Player~0, in terms of the sequence
$\Omega( v_0 )\Omega( v_1 ) \dots$ of priorities
appearing in the play. Thus, a winning condition is 
a set $W\subseteq C^\omega$ of infinite
sequences of priorities.

A \emph{(deterministic) strategy} for Player~$\sigma$ is a 
partial function $f:V^*V_\sigma\ra V$
that assigns to finite paths through $\Gg$ ending in a position
$v\in V_\sigma$ a successor $w\in vE$. A play $v_0v_1\dots \in V^\omega$
is \emph{consistent} with $f$ if, for each initial
segment $v_0\dots v_i$ with $v_i\in V_\sigma$, we have 
that $v_{i+1}=f(v_0\dots v_i)$.
We say that such a strategy $f$
is winning from position $v_0$ if every play that starts at $v_0$ and
that is consistent with $f$ is won by
Player~$\sigma$.  The \emph{winning region} of Player~$\sigma$,
denoted $W_\sigma$, is the set of positions from which Player~$\sigma$
has a winning strategy.  A game $\Gg$ is \emph{determined} if $W_0\cup
W_1=V$, i.e., if from each position one of the two players has a
winning strategy.

Winning strategies can be rather complicated.  Of special interest are
simple strategies, in particular \emph{finite memory strategies} and
\emph{positional strategies}. While positional strategies only depend
on the current position, not on the history of
the play, finite memory strategies have access to bounded amount of
information on the past.  Finite memory strategies can be defined as
strategies that are realisable by finite automata.

More formally, a strategy with memory $M$ for Player~$\sigma$ is given
by a triple $(m_0,U,F)$ with initial memory state $m_0\in M$,
a memory update function $U: M\times V\ra M$
and a next-move function $F: V_\sigma\times M\ra V$.
Initially, the memory is in state $m_0$
and after the play has gone through
the sequence $v_0v_1\ldots v_{m}$
the memory state is $u(v_0\dots v_{m})$,
defined inductively by $u(v_0\dots v_{m} v_{m+1})= U(u(v_0\dots 
v_{m}),v_{m+1})$.
In case $v_{m}\in V_\sigma$, the next move from
$v_1\dots v_{m}$, according to the
strategy, leads to $F(v_{m},u(v_0\dots,v_m))$.
In case $M=\{m_0\}$, the strategy is positional;
it can be described by a function $F: V_\sigma\ra V$.

\defn A game is \emph{positionally determined}, if it is determined,
and each player has a positional
winning strategy on his winning region.  \edefn

Clearly, if the arena is a forest, then all strategies
are positional, so the game is positionally determined
if, and only if, it is determined.

Throughout the paper, we assume the Axiom of Choice.

\subsection{Games with infinitely many priorities}

In the context of finite-memory determinacy or
positional determinacy of infinite games it is usually assumed that the
range of the priority function is finite, and the winning condition is
defined by a formula on infinite paths (from S1S or LTL, say)
referring to the predicates $(P_c)_{c\in C}$, or by an
automata-theoretic condition like a Muller, Rabin, Streett, or parity
(Mostowski) condition (see e.g.
\cite{GraedelThoWil02,DziembowskiJurWal97,Zielonka98}).  In Muller
games the winner of a play depends only on the set of priorities that
have been seen infinitely often; it has been proved by Gurevich and
Harrington \cite{GurevichHar82}
that Muller games are determined and that the winner has a
finite-memory winning strategy.  An important special
case of Muller games are \emph{parity games} where the least (or
greatest) priority occurring infinitely often determines the winner.

Here we will extend the study of
positional determinacy  to games with
infinitely many priorities. Specifically
we are interested in games with priority
assignments $\Omega:V\ra\omega$. 
Besides the obvious theoretical interest,
such games arise in several areas. 
For instance, the winning conditions of pushdown games  are specific
instances of abstract winning conditions in games
with infinitely many priorities. It is interesting to 
study these games in a general setting, and to
isolate the winning conditions that lead to
positional determinacy on arbitrary arenas, not
just on specific ones like pushdown games.

Based on priority assigments $\Omega: V\ra\omega$
we will first consider the following classes of games.

\begin{description}
\item[Infinity games] are games where Player~0 wins
precisely those infinite plays in which no
priority appears infinitely often.
\item[Parity games] are games where Player~0
wins the infinite plays where the least
priority seen infinitely often is even,
or where all priorities appear only finitely
often.
\item[Max-parity games] are games where Player~0
wins if the maximal priority occurring infinitely
often is even, or does not exist.
\end{description}

Note that we have chosen the definitions so that in case no priority
appears infinitely often, the winner is always Player~0.  It is clear
that these games are determined, because the winning conditions are
Borel sets, and a fundamental result due to Martin \cite{Martin75}
states that all Borel games are determined.  To be more precise, the
infinity and parity winning conditions are on the ${\boldsymbol
\Pi^0_3}$-level of the Borel hierarchy. Indeed,
note that for any $m\in\omega$ the set $A_m$ of words that contain 
infinitely many occurences of $m$ is in ${\boldsymbol \Pi^0_2}$
since it is the countable intersection of the open
sets $A_m^n:=(\omega^* m)^n\omega^\omega$, for all $n\in\omega$.
Now the parity condition can be expressed as the
the set of infinite words $x=x_0x_1x_2\dots$
such that for all odd $m$, either $x\not\in A_m$ or there
is an even number $k<m$ such that $x\in A_k$. 
Similarly, it is easy to see that the max-parity condition
is on the ${\boldsymbol \Delta^0_4}$-level of the Borel hierarchy.

For games with only finitely many priorities,
min-parity and max-parity winning conditions
can be (and are) used interchangeably.
This is not the case when we have
infinitely many priorities.

\proposition \label{max-parity}Max-parity games with infinitely many priorities
in general do not admit finite memory winning strategies.
\eprop

\look Consider the max-parity game with
positions $V_0=\{0\}$ and $V_1= \{2n+1: n\in\N\}$
(where the name of a position is also its priority),
such that Player 0 can move from $0$ to
any position $2n+1$ and Player~1 can move
back from $2n+1$ to $0$. Clearly Player~0 has
a winning strategy from each position
but no winning stategy with finite memory.
\hx

However, we will see that (min-)parity games
with priorities in $\omega$
are positionally determined.

\subsection{Strategy forests}

Let $f$ be a strategy for Player~$\sigma$
in the game $\Gg=(V,V_0,V_1,E,\Omega)$. For
any initial position $v_0$ of the game,
we can associate with $f$ the \emph{strategy tree}
$\Tt_f$, the tree of all
plays that start at $v_0$ and that are consistent
with $f$. In the obvious way,
$\Tt_f$ can itself be considered as a game graph,
with a canonical homomorphism $h:\Tt_f\ra \Gg$.
For every position $v$ of $\Gg$, we call
the nodes $s\in h^{-1}(v)$ the
\emph{occurrences of $v$} in $\Tt_f$.
Since we assume that strategies are
deterministic
every occurence of a node $v\in V_\sigma$
has precisely one successor in the strategy forest $\Tt_f$,
whereas every occurrence of a node $v\in V_1$
has precisely as many successors in $\Tt_f$
as $v$ has in $\Gg$.
If $f$ is a winning strategy from $v_0$,
then every path through $\Tt_f$
is a winning play for Player~$\sigma$.
If we consider a set of initial
positions (like the entire winning region $W_\sigma$)
then $\Tt_f$ is a \emph{strategy forest}
with a separate tree for each initial position.

\medskip By moving from game graphs to strategy forests
we can eliminate the interaction
between the two players and thus simplify the
analysis. We already know that the games that we study
are determined.
To prove positional determinacy
we proceed as follows.

We take a winning strategy and
define a collection of
well-founded pre-orders on its strategy forest.
We then define a positional winning strategy
for the original game, by copying
for each position in the winning region,
the winning stategy from a minimal occurrence
of the position in the strategy tree.
We then show that the resulting positional
strategy is indeed winning.

To simplify the exposition we first discuss infinity games.  Note that
these can be seen as a special case of parity games.
Indeed, if we change the priorities of an infinity
game $\Gg$ so that all priorities become odd, by setting
$\Omega'(v):=2\Omega(v)+1$, and replace the infinity winning condition
by the parity condition, then the resulting parity game $\Gg'$ is
equivalent to $\Gg$.

\section{Infinity Games}

We start with some remarks on
arbitrary transition systems. We will then
apply them to strategy forests.

Given any transition system $\Kk=(S,E,P)$ with
set of states $S$, transition relation $E$
and atomic proposition $P$, we
assign to each state $s$ an ordinal $\alpha(s)$
or $\infty$.
Informally, $\alpha(s)$ tells us how
often a path from $s$ can hit $P$.
To define this precisely, we
proceed inductively.
For any ordinal $\alpha$, let
$X^\alpha$ be the set of all $s\in S$
such that whenever a path from $s$
hits a node $t\in P$, then all successors
of $t$ belong to $\bigcup_{\beta<\alpha} X^\beta$.
Finally, let $\alpha(s)=\min\{\alpha: s\in X^\alpha\}$.
If $s$ is not contained in any $X^\alpha$, the we put
$\alpha(s)=\infty$.

\usec{Remark. }We can equivalently define $\alpha(s)$ in terms
of closure ordinals in the modal $\mu$-calculus.
Consider the formula
$\mu X. \phi(X)$, with $\phi(X):= \nu Y . (P\ra \Box X)\land \Box Y$.
It expresses that on all paths, there are
only finitely many occurrences of $P$.
We define the stage $X^\alpha$ of the least fixed point
induction via $\phi(X)$ by
$X^\alpha = \{ s: \Kk,s\models \phi(X^{<\alpha})\}$
where $X^{<\alpha}:=\bigcup_{\beta<\alpha} X^\beta$.
It is easily seen that this coincides with the definition given
above.

\usec{Remark. }Note that although
$\mu X.\phi(X)$ expresses that on every path there
are only finitely many occurrences of $P$ the closure
ordinals need not be finite. For a simple example,
consider an infinite path $v_0v_1v_2\ldots$
without occurences of $P$ and attach to each $v_n$
another infinite path on which $P$
is seen precisely $n$ times. On all these attached
paths, $\alpha(s)$ will take only finite values,
but $\alpha(v_n)=\omega$ for all $n$.

The following lemma is a direct consequence of the
definitions. 

\lemma \label{lemma:sig}
Suppose that every path in $\Kk$ contains only
finitely many occurrences of $P$
(i.e., $\Kk,s\models \mu X.\phi(X)$ for all $s$).
Then $\alpha(s)\geq \alpha(t)$ for all edges $(s,t)$ of $\Kk$,
and the inequality is strict for $s\in P$.
\elem

Assume next that we have a transition system
$\Kk=(S,E,P_0,P_1,P_2,\ldots)$
with infinitely many atomic propositions
$P_n$.
Proceeding as above for $P_n$ instead of $P$,
we obtain, for each $n$, a function $\alpha_n$ mapping
states $s\in S$ to ordinals.
The \emph{signature} of $s$
is $\sig(s):=\langle \alpha_n(s): n <\omega\rangle$;
we compare signatures lexicographically.
Further, for each $n<\omega$,
let $\sig_n(s)= \langle \alpha_0(s),\ldots,\alpha_n(s)\rangle$
and let $s <_n t$ denote that
$\sig_n(s) < \sig_n(t)$ (i.e., that the signature
of $s$ is strictly smaller than the signature of $t$ on the first
$n+1$ positions).
Similarly, let $s \leq_n t$ denote that $\sig_n(s)\leq \sig_n(t)$.

Note that $s <_{n} t$ implies $ s <_{n+1} t$ and that each
pre-order $<_n$ is well-founded 
(i.e., all descending chains are finite).  On the other side,
when we have infinitely many $P_n$, the lexicographic order of
unrestricted signatures admits infinite descending chains.

\theorem \label{thm:infinity}
Infinity games are positionally determined
\etheo

\look Let $W_0$ and $W_1$ be the winning regions
of the two players for the infinity game on
the arena $\Gg$. Note that the situation for the two
players is not symmetric, so we have to consider them
separately.

Let $f$ be any winning strategy for Player~0 on $W_0$.  If $f$ is
positional then we are done.  Otherwise, we consider the strategy
forest $\Tt_f$ and the canonical homomorphism $h:\Tt_f\ra \Gg$.  In
$\Tt_f$ every path is winning for Player~0, and thus hits each $P_n$
only finitely often.  Hence the functions $\alpha_n(s)$ are defined
and satisfy the properties of Lemma~\ref{lemma:sig}.

We define a positional strategy $f'$ for Player~0 as follows.  Select
a function $s :W_0\ra \Tt_f$ that associates with each vertex $v\in
W_0$ of priority $n$ a $<_n$-minimal element $s(v)\in h^{-1}(v)$
(i.e., a $<_n$-minimal occurrence of $v$ in the strategy forest).  If
$v$, and hence also $s(v)$, is a node of Player~0, then there is a
unique successor $t$ of $s(v)$ in $\Tt_f$; define $f'(v):= h(t)$.
Further, we define values of $\alpha_n$ (and hence $\sig_n$) on $W_0$
by $\alpha_n(v):= \alpha_n(s(v))$.

We claim that $f'$ is winning from each node
$v_0\in W_0$. Otherwise there exists a play
$v_0v_1v_2\ldots$ that is consistent with $f'$
and winning for Player~1.
Let $n$ be the least priority seen infinitely
often on this play; take a suffix
of the play on which
priorities smaller than $n$ do
no longer occur. We claim that the values
of $\sig_n$ never increase on this suffix.

To see this, consider a move from $v$ to $w$ in
this suffix
and the corresponding moves in $\Tt_f$
from $s:=s(v)$ to $t$ with $h(t)=w$.
By construction, and since $v$ and $w$ have
priorities $\geq n$, we have
$\sig_n(v) = \sig_n(s)$ and $\sig_n(w) \leq \sig_n(t)$.
By Lemma~\ref{lemma:sig}, we have $\alpha_m(s)\geq \alpha_m(t)$
for all $m$ and the inequality is strict if $m$
is the priority of $s$.
It follows that
\[ \sig_n(v)=\sig_n(s)\geq \sig_n(t)\geq \sig_n(w)\]
and $\sig_n(v)>\sig_n(w)$ in case $v\in P_n$.
Since there are infinitely many
nodes $v_{i_1}, v_{i_2},\ldots$ of priority $n$ in the suffix,
we obtain an infinite descending chain
\[ \sig_n(v_{i_1}) > \sig_n(v_{i_2}) > \cdots \]
which is impossible.
Hence $f'$ is indeed a winning strategy.

\medskip We now consider the case of Player~1.  Let $g$ be a strategy
for Player~1 on $W_1$, with strategy forest $\Tt_g$ and canonical
homomorphism $h:\Tt_g\ra \Gg$.  We define the \emph{$0$-ancestor} of a
node $s\in \Tt_g$ to be the closest ancestor of $s$ that has priority
$0$. Note that $0$-ancestors may be
undefined.  More generally, the \emph{$m$-ancestor} of $s$ is the
closest ancestor of priority $m$, provided it lies between $s$ and the
$j$-ancestor of $s$, for all $j<m$ for which the $j$-ancestor is defined.
We can thus associate with every node $s$ of priority $m$ an
$(m+1)$-tuple $a(s)=\langle a_0(s),\ldots,a_m(s)\rangle\in
(\Tt_g\cup\{\bot\})^{m+1}$ of ancestors, where $a_i(s)=\bot$ means
that the $i$-th ancestor of $s$ is not defined. Observe that $a_m(s)=s$
as $s$ is an ancestor of itself.

We fix a well-order $<$ on $\Tt_g\cup\{\bot\}$ (with
maximal element $\bot$) and we compare tuples of
ancestors via the lexicographical order that is induced by $<$.  We
can then associate with every $v\in W_1$ of priority $m$ the node
$s(v)\in h^{-1}(v)$ with the minimal tuple of ancestors.  Note that $s(v)$
is well-defined, because every position $v\in W_1$ has at least one
occurrence $s\in \Tt_g$, and $a_m(s)=s$ if $m$ has priority $m$ (so at
least one ancestor is defined).  We extend the ancestor function to
$W_1$ by setting $a(v):= a(s(v))$; this assigns to every node in $W_1$
a tuple of ancestors in $\Tt_g$.  To define the positional strategy
$g'$, we select for any $v\in V_1\cap W_1$ the unique successor $t$ of
$s(v)$ and set $g'(v):=h(t)$.

We claim that this strategy is winning for Player~1.  Suppose
conversely that there is a losing play respecting the strategy. Then
no priority appears infinitely often on this play. Consider the suffix
of the play after the last appearance of priority $0$.  Let us look at
the $0$-ancestors of the positions in this suffix.  These ancestors
can only get smaller as the play proceeds.  Indeed a move from $v$ to
$w$ in such a play corresponds to a move from $s(v)$ to $t$ in $\Tt_g$
with $h(t)=w$. Since $w$ does not have priority 0, $a_0(s) = a_0(t)$
and therefore $a_0(v)=a_0(s) = a_0(t)\geq a_0(w)$.  This means that
from some moment on all positions in the play will have the same
$0$-ancestor.  Consider the suffix of the play consisting only of
these vertices.  Next do the same with priority $1$. We find a
position after which the $1$-ancestor stabilises. Observe that it is a
descendant of the $0$-ancestor and that there is no occurrence of
priority $0$ on the path between the two.  Proceeding in this way we
construct a path in the strategy tree $\Tt_g$ on which no priority
appears infinitely often. But this is impossible, since $g$ was a
winning strategy for Player~1.  \hx

\section{Parity games}

For parity games we proceed quite similarly,
but we have to consider more complicated
orderings on the strategy forests.

Consider a transition system
$\Kk=(S,E,P,Q)$ with two atomic propositions
$P$ and $Q$. We
assign to each state $s$ an ordinal $\beta(s)$
or $\infty$ which,
informally, tells us how
often a path from $s$ can hit $P$
before seeing $Q$.
Let $X^0$ be the set of all $s$ such that all
paths from $s$
hit $Q$ before hitting $P$,
and for $\beta>0$, let
$X^\beta$ be the set of all $s$
such that whenever a path from $s$
hits a node $t\in P$, then all successors
of $t$ belong to $X^{<\beta}$.
Finally, let $\beta(s)=\min\{\beta: s\in X^\beta\}$.

\medskip Again, we have an equivalent definition in terms of the modal
$\mu$-calculus.  This time, consider the formula $\mu X. \phi(X)$,
with $\phi(X):= \nu Y . (\neg P\lor \Box X)\land (Q\lor \Box Y)$.  It
expresses that on all paths, there are only finitely many occurrences
of $P$ before seeing $Q$.  Then $\beta(s)$ is the stage at which the
least fixed point induction defined by $\phi(X)$ becomes
true at node $s$.

\lemma \label{lemma:sig2}
Suppose that every path in $\Kk$ contains only
finitely many occurrences of $P$ before hitting $Q$.
Then $\beta(s)\geq \beta(t)$ for all edges $(s,t)$ of $\Kk$ with $s\not\in Q$,
and the inequality is strict for $s\in P$.
\elem

For infinity games we have defined ordinals
$\alpha_n(s)$ telling us
how often a path from $s$ can see priority $n$,
independently for each $n$.
Now we need different bounds $\beta_n$
which, informally, describe how often
a path can hit the odd priority $n$ before
seeing a smaller one.

Let $\Gg$ be a parity game, and
let $\Tt_f=(S,E,P_0,P_1,P_2,\ldots)$
be the strategy forest of a winning strategy $f$
for Player~0. Note that for every odd priority $n$,
each path through $\Tt_f$
sees only finitely many occurrences of $n$
before seeing a priority $<n$.
Hence,
proceeding as above for $P:=P_n$
and $Q:=\bigcup_{m<n} P_m$
we obtain, for each odd $n$, a function $\beta_n$ mapping
nodes $s\in\Tt_f$ to ordinals.
The \emph{0-signatures} of $s$
are $\sig^0_n(s):=\langle \beta_1(s),\beta_3(s)\ldots,\beta_{n'}(s)\rangle$,
where $n'=n$ for odd $n$ and $n'=n-1$ for even $n$;
let $s <^0_n t$ denote that
$\sig^0_n(s) < \sig^0_n(t)$. Further, $s\leq^0_n t$ means
that $\sig^0_n(s) \leq \sig^0_n(t)$.

For strategy forests $\Tt_g$ of winning strategies
of Player~1, we proceed dually,
associating with every node $s$
ordinals $\beta_n(s)$, for even $n$.
We then define $1$-signatures
$\sig^1_n(s)=\langle\beta_0(s),\beta_2(s),\ldots,\beta_{n'}(s)\rangle$
(where $n'$ is the largest even number not exceeding $n$)
and the corresponding signature orderings $<^1_n$.

Again, we immediatley see that $s <^i_n t$ implies
$s <^i_{n+1} t$ and that each $\leq^i_n$ is a well-founded.
Further, these orderings have very useful properties
on strategy forests.

\lemma \label{lemma:sig3} Let $\Tt_f$ be the strategy forest
associated with a winning strategy for Player~0 for a parity game.
Then $t \leq^0_{\Omega(s)} s$ for all edges $(s,t)$ of $\Tt_f$ and the
inequality is strict if $\Omega(s)$ is odd.  In a strategy
forest $\Tt_g$ of Player~1, we have $t
\leq^1_{\Omega(s)} s$ for all edges $(s,t)$ and the inequality is
strict if $\Omega(s)$ is even.  \elem

\look If $(s,t)$ is an edge in $\Tt_f$, then by Lemma~\ref{lemma:sig2},
$\beta_m(t) \leq \beta_m(s)$ for $m\leq \Omega(s)$,
and, if $n=\Omega(s)$ is odd, and $\beta_n(t) < \beta_n(s)$.
Similarly for $\Tt_g$.
\hx

\theorem \label{thm:parity}
Parity games with priorities in $\omega$
are positionally determined.
\etheo

\look The proof for Player~0 is precisely the same as
for infinity games, using $0$-signatures
and the associated orderings $<^0_n$.

For Player~1 we combine the approach for infinity games based on
ancestors in the strategy forest with comparisons based on
$1$-signatures.  As in the proof of Theorem~\ref{thm:infinity} we
associate with every node $s$ of priority $m$ in the strategy tree
$\Tt_g$ the $(m+1)$-tuple 
$a(s)=\langle a_0(s),\ldots,a_m(s)\rangle\in (\Tt _g\cup\{\bot\})^{m+1}$  
of ancestors.  
 
For  each $i\in \omega$ we fix a well-order $\lhd_i$
extending $<^1_i$. Moreover we assume that $\bot$ is bigger in 
the $\lhd_i$-order than all the nodes. Let $s$, $s'$ be two nodes of
$\Tt_g$ of the same priority $m$.  We write $s\prec_m s'$ if there is
$i\leq m$ such that $a_i(s)\lhd_i a_i(s')$ and $a_j(s)=a_j(s')$ for all
$j<i$. Observe that $\prec_m$ is a well order on vertices of priority $m$.


For any position $v\in W_1$ of priority $m$ we now
take the $\prec_m$-minimal occurrence $s(v)$ in $\Tt_g$ and define
ancestors by $a(v):= a(s(v))$.  For $v\in V_1\cap W_1$ we consider the
unique successor $t$ of $s(v)$ and set $g'(v):=h(t)$.  This defines a
positional strategy $g'$ for Player~1 on $\Gg$.

We claim that this strategy is winning on $W_1$.
Suppose conversely that there is a losing play respecting the
strategy.
Then either no
priority appears infinitely often on this play, or the smallest
priority occurring infinitely often is even.

If no priority occurs infinitely often, then
we can
proceed as in the proof of Theorem~\ref{thm:infinity}
to show that all ancestors eventually stabilise
on the play, and thus obtain an infinite path in $\Tt_g$ on which
no priority appears infinitely often. This is impossible
since $g$ is a winning strategy for Player~1.
If the minimal priority $p$ appearing infinitely often is even, then
we consider a suffix of the play that contains only priorities $\geq
p$.  By the same reasoning as in the first case it follows that all
$q$-ancestors, for $q<p$, eventually stabilise on the play. Consider
the suffix of the play after this has happened.  A move from $v$ to
$w$ on this suffix corresponds to a move from $s(v)$ to $t$ in
$\Tt_g$.  By definition, the $p$-ancestor of $v$ is $a_p(v)=a_p(s(v))$
and $a_p(w)\preceq_p a_p(t)$.  On $\Tt_g$ we obviously have $a_p(t)=t$
if $\Omega(t)=p$ and $a_p(t)=a_p(s(v))$ if $\Omega(t)>p$.  Now $t$ is
a descendant of $a_p(s)$, so by Lemma~\ref{lemma:sig3} we have
$t < ^1_p a_p(s)$; for the case that
$\Omega(t)=p$ this means that $a_p(t) =t <^1_p a_p(s)$.  Since
$\prec_p$ extends $<^1_p$ on nodes of priority $p$, we have that
$a_p(w) \preceq_p a_p(v)$ and that the inequality is strict if
$\Omega(w)=p$.
But on the suffix we have an inifinite sequence
of positions with priority $p$,
and hence an infinite $\prec_p$-decreasing chain of
$p$-ancestors, which is impossible.
\hx

\noindent{\bf Remark: Parity games over larger ordinals. }
We can also define parity games
with a priority function $\Omega: V\ra \alpha$
taking values in a larger set of ordinals than $\omega$.
Recall that any ordinal can be written in a unique way
as a sum $\lambda+n$ where $\lambda$ is a limit ordinal 
and $n<\omega$. We call $\lambda+n$ even if $n$ is.
The question arises whether the positional determinacy
of parity games over $\omega$ extends to
larger ordinals. However, a tiny modification
of the game in Proposition~\ref{max-parity} shows that 
this is not the case. Indeed, if we replace in
that game priority 0 by $\omega$, and use the (min-)parity
winning condition, then Player~0 has
a winning strategy from each position
but no winning strategy with finite memory.
For larger ordinals, a similar construction applies.
This proves that parity games over ordinals $\alpha>\omega$
in general do not guarantee finite memory winning strategies.

Essentially the same construction shows that
finite-memory determinacy also fails for some other
variants of parity games over $\omega$, such as
\begin{itemize}
\item parity games where the priority function is partial
(i.e., not all vertices have a priority),
\item parity games with priorities on edges rather than vertices.
\end{itemize}

\section{Muller games}

Why do parity games and max parity games behave differently?  Both are
Muller conditions (i.e. they refer only to the set of priorities seen
infinitely often) and the question arises which properties of Muller
conditions are responsible for positional determinacy or determinacy
with finite memory.  In this section we assume that
the set of priorities is countable.  This is reasonable as on each
play one can see only a countable number of them.

\defn A Muller condition over a set $C$ of
priorities is written in the form $(\Ff_0,\Ff_1)$ where
$\Ff_0\incl\Pp(C)$ and $\Ff_1=\Pp(C)-\Ff_0$.  A play in a game with
Muller winning condition $(\Ff_0,\Ff_1)$ is won by Player~$\sigma$ if,
and only if, the set of priorities seen infinitely often
in the play belongs to $\Ff_\sigma$.  \edefn

For infinity games, we have $\Ff_0=\{\emptyset\}$
and $\Ff_1=\Pp(\omega)-\{\emptyset\}$. For parity games,
\begin{align*}
\Ff_0 &= \{X\subseteq \omega: \min(X) \text{ is even}\}\cup \{\emptyset\}\\
\Ff_1 &= \{X\subseteq\omega: \min(X) \text{ is odd}\}
\end{align*}
For max-parity games, we have
\begin{align*}
\Ff_0 &= \{X\subseteq \omega: \text{ if $X$ is finite and non-empty,}\\ 
&\qquad\text{then $\max(X)$ is even}\}\\
\Ff_1 &= \{X\subseteq\omega: X \text{ is finite, non-empty, and}\\ 
&\qquad\text{$\max(X)$ is odd}\}
\end{align*}

\defn We say that $(\Ff_0,\Ff_1)$ \emph{guarantees positional winning 
strategies} if \emph{all}
games with winning condition $(\Ff_0,\Ff_1)$ are positionally determined.
\edefn

Following McNaughton \cite{McNaughton93} and Zielonka
\cite{Zielonka98} we say that $\Ff_\sigma$  
has a \emph{strong split} 
if there exist sets  $X_0,X_1\in \Ff_\sigma$ with
$X_0\cap X_1\not=\emptyset$ and 
$X_0\cup X_1\in \Ff_{1-\sigma}$. 
Zielonka \cite{Zielonka98} has shown that a Muller condition over a
\emph{finite} set of priorities guarantees positional 
winning strategies if, and only if,

\begin{description}
\item[(P0)] $\Ff_0$ and $\Ff_1$ have no
\emph{strong splits}.
\end{description}

\usec{Remark. } A weak split is a pair of \emph{disjoint}
sets with $X_0,X_1\in \Ff_\sigma$ and
$X_0\cup X_1\in \Ff_{1-\sigma}$. Muller conditions over finite
sets of priorities may have weak splits and still guarantee
positional winning strategies.
The simplest case is when $\Ff_0$ consists of 
the set $\{0,1\}$, but $\{0\}$ and $\{1\}$
belong to $\Ff_1$.

\medskip
We want to find a similar characterisation
of Muller conditions with positional winning strategies
for the case of infinite sets of priorities.

We observe that for infinity games and parity games
$\Ff_0$ and $\Ff_1$ are closed under unions
and  non-empty intersections of chains:

\begin{description}
\item[(P1)] For every infinite descending chain $X_1\supseteq X_2 \supseteq
  \dots$ of elements of $\Ff_\sigma$
  either $\bigcap_{i<\omega} X_i=\emptyset$ or it is an element of $\Ff_\sigma$.
\item[(P2)] For every chain  $X_1\subseteq X_2 \subseteq
\dots$ of elements of $\Ff_\sigma$,
also $\bigcup_{i<\omega} X_i$ belongs to $\Ff_\sigma$.
\end{description}

On the other side, for the max-parity condition,
$\Ff_0$ is not closed under non-empty intersections of chains
(take $X_i = \{1\}\cup\{ n: n>i\}$) and $\Ff_1$ is not
closed under unions of chains (take $X_i= \{j: j\leq 2i+1\}$).
Condition (P1) fails also for min-parity condition for ordinals
$\alpha>\omega$. Indeed we have  
$\Ff_1=\{X\subseteq \alpha: \min(X) \text{ is odd}\}$
which is not closed under non-empty intersections of chains
(take $X_i=\{\omega\}\cup\{n: 2i+1\leq n <\omega\}$).

We will show first, that condition (P1) 
is necessary for the positional determinacy of
a Muller condition.

\lemma\label{lem:P1}
If there is an infinite sequence $X_1\supseteq X_2 \supseteq \cdots$
of elements of $\Ff_{1-\sigma}$ with $\bigcap X_i =Y\not=\emptyset$ and $Y\in
\Ff_\sigma$ then there is game with winning condition $(\Ff_0,\Ff_1)$
that Player~$\sigma$ wins, but needs infinite memory to do
so.
\elemma

\look Consider the following game where circles denote positions
of Player~$\sigma$ and boxes positions of Player$~(1-\sigma)$.

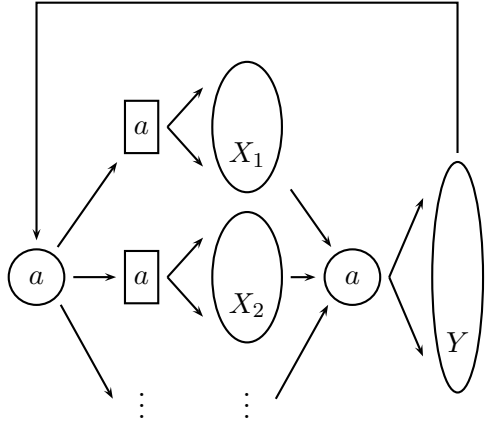
\begin{center}
\psset{xunit=1.4cm,yunit=2cm,arcangle=15}
\begin{pspicture}(4,3)
\rput(0,1){\mcnode{c}{$a$}}
\rput(1,2){\mbnode{di}{$a$}}
\rput(1,1){\mbnode{dii}{$a$}}
\rput(1,.2){\rnode{dotsi}{$\vdots$}}
\rput(2,2){\rnode{ei}{\psovalbox{\makebox{\rule{0pt}{1cm}$X_1$}}}}
\rput(2,1){\rnode{eii}{\psovalbox{\makebox{\rule{0pt}{1cm}$X_2$}}}}
\rput(2,.2){\rnode{dotsii}{$\vdots$}}
\rput(3,1){\mcnode{f}{$a$}}
\rput(4,1){\rnode{g}{\psovalbox{\makebox{\rule{0pt}{2cm}$Y$}}}}
\ncline[offsetB=30pt]{a}{b}
\ncline[offsetB=-30pt]{a}{b}
\ncline{b}{c}
\ncline{c}{di}
\ncline{c}{dii}
\ncline[offsetB=-.3cm]{c}{dotsi}
\ncline[offsetB=15pt]{di}{ei} \ncline[offsetB=-15pt]{di}{ei}
\ncline[offsetB=15pt]{dii}{eii} \ncline[offsetB=-15pt]{dii}{eii}
\ncline{ei}{f}
\ncline{eii}{f}
\ncline[offsetA=-.3cm]{dotsii}{f}
\ncline[offsetB=30pt]{f}{g}
\ncline[offsetB=-30pt]{f}{g}
\ncbar[angleA=90,angleB=90,armA=2cm,armB=3cm]{g}{c}
  \end{pspicture}
\end{center}

\medskip
Here $a$ is some arbitrary element of
$Y$.  A play in this game is an infinite sequence of 
subplays; in each subplay Player~$\sigma$ first decides 
from which $X_i$ the opponent is going
to choose next. After Player~$(1-\sigma)$ has made his choice,
Player~$\sigma$ can select an element from $Y$.

If Player~$\sigma$ allows her opponent to choose from some $X_i$ infinitely
often then Player~$(1-\sigma)$ can make all elements of $X_i$ appear infinitely
often on the play. This means that in order not to lose, Player~$\sigma$
must permit Player~$(1-\sigma)$ to choose from each $X_i$ only finitely
often. If she does this then she wins as she can make sure
that each element of $Y$ is seen infinitely often thanks to the last 
part of the each subplay.
Thus Player~$\sigma$ has a winning strategy, but none
that uses only finite memory.
\hx

To show the necessity of conditions (P0) and (P2) we consider the
following game. 

\vspace{5mm}
\centerline{\psset{xunit=1.5cm,yunit=1.5cm,arcangle=15}
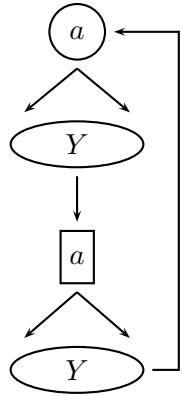
\begin{pspicture}(2,3)
  \rput(1,3){\mcnode{a}{$a$}} 
  \rput(1,2){\ovalnode{b}{\quad$Y$\quad}}
  \rput(1,1){\mbnode{c}{$a$}} 
  \rput(1,0){\ovalnode{d}{\quad$Y$\quad}} 
  \ncline[offsetB=20pt]{a}{b} 
  \ncline[offsetB=-20pt]{a}{b}
  \ncline{b}{c}
  \ncline[offsetB=20pt]{c}{d} 
  \ncline[offsetB=-20pt]{c}{d}
  \ncbar[angleA=0,angleB=0]{d}{a}
  \end{pspicture}
} \vspace{2ex}

\medskip
Here, $Y$ is a set and $a\in Y$.
The arrows to the ovals with $Y$
mean that the player can choose any element of $Y$.  
Clearly, if $Y\in \Ff_\sigma$, then
Player~$\sigma$ can win by visiting
all elements of $Y$ infinitely often. 
However, if Player~$\sigma$ plays
memoryless
then she must select a fixed element
$b$, and her opponent can chose
an arbitrary set $X\subseteq Y$ of nodes
and make sure that the set of nodes visited infinitely 
often is $\{a,b\}\cup X$.
More generally, if Player~$\sigma$ plays
with a finite memory strategy
this amounts to selecting a finite set 
$B$; Player~$(1-\sigma)$ can then win if there
exists a set $X\in\Ff_{1-\sigma}$ with
$\{a\}\cup B \subseteq X\subseteq Y$.

\lemma If $\Ff_{1-\sigma}$ contains a strong split,
then there is a game with winning condition
$(\Ff_0,\Ff_1)$ that is won by Player~$\sigma$,
but not with a positional strategy.
\elemma

\look Let $X_0\cup X_1\in\Ff_\sigma$
with $X_0,X_1\in\Ff_{1-\sigma}$ and
$X_0\cap X_1\neq \emptyset$.
Take the game above with $Y=X_0\cup X_1$
and $a\in X_0\cap X_1$.
Player~$\sigma$ wins since $Y=X_0\cup X_1\in\Ff_\sigma$.
However, she cannot win
positionally. Indeed the single element $b$ selected
by a positional strategy of Player~$\sigma$
belongs to $X_i$ ($i=0$ or $1$), and
Player~$1-\sigma$ can win by making sure
that all elements of $X_i$, and only these,
are visited infinitely often.
\hx

\lemma If $\Ff_{1-\sigma}$ is not closed under
unions of chains, then there is a game
with winning condition
$(\Ff_0,\Ff_1)$ that Player~$\sigma$ wins,
but needs infinite memory to do so.
\elemma

\look Let $X_1\subseteq X_2\subseteq\cdots$
be an infinite ascending chain in $\Ff_{1-\sigma}$ with
$\bigcup_i X_i \in\Ff_\sigma$.
Take the game described above
with $Y=\bigcup_i X_i$ and $a\in X_1$.
Again Player~$\sigma$ wins since $Y\in\Ff_\sigma$.
But if Player~$\sigma$ plays with finite memory,
this amounts to selecting a finite set $B\subseteq Y$
of elements that she visits infinitely often.
Since $B$ is finite $B\subseteq X_i$ for some
$i$; hence Player~$1-\sigma$ can make sure that
the set of elements visited infinitely often
is $X_i$ and wins.
\hx

In the remaining part of the section we will 
characterise the Muller conditions satisfying 
(P0), (P1) and (P2) in a different way, via Zielonka paths,
and then show that any such condition can be reformulated
as a parity condition 
over an ordinal $\alpha\leq\omega$.
In particular, this implies that these closure properties
are necessary and sufficient to guarantee positional
determinacy on all game graphs.

\defn The \emph{Zielonka tree} of a Muller condition $(\Ff_0,\Ff_1)$
over $C$ is a tree $Z(\Ff_0,\Ff_1)$ whose nodes are labelled with
pairs $(X,\sigma)$ such that $X\in\Ff_\sigma$. Let
$\sigma$ be the player that wins with the set of
all priorities, i.e. $C\in
\Ff_{\sigma}$ with $C=\bigcup \Ff_0\cup\bigcup\Ff_1$.  The Zielonka
tree $Z(\Ff_0,\Ff_1)$ exists, if for every maximal $Y\in
\Ff_{1-\sigma}$ the Zielonka tree $Z(\Ff_0\cap\Pp(Y),\Ff_1\cap
\Pp(Y))$ exists and every set in $\Ff_{1-\sigma}$ is a subset of some
maximal set in $\Ff_{1-\sigma}$.  In that case $Z(\Ff_0,\Ff_1)$
consists of a root, labeled by $(C,\sigma)$, to which we attach as
subtrees the Zielonka trees $Z(\Ff_0\cap\Pp(Y),\Ff_1\cap \Pp(Y))$, for
the maximal sets $Y\in \Ff_{1-\sigma}$.  (In particular, if
$\Ff_{1-\sigma}=\emptyset$, then the Zielonka tree consists of a
single node.)  \edefn

For Muller conditions over a finite set $C$, the Zielonka
tree always exists, and it is a fundamental tool
for analysing the memory that is required for solving
Muller games \cite{DziembowskiJurWal97}. 
For infinite sets $C$, the Zielonka tree
need not exist, since there is no guarantee,
that for $X\in\Ff_\sigma$, the set $\Pp(X)\cup\Ff_{1-\sigma}$
contains maximal elements. For instance the max-parity
condition does not have a Zielonka tree.

\proposition\label{prop:max} For every Muller condition
satisfying property {\rm (P2)} the Zielonka tree exists. 
\eprop

\look By (P2) the union over any chain $Y_0\incl Y_1\incl \dots$ in
$\Pp(X)\cap\Ff_\sigma$ is again contained $\Pp(X)\cap\Ff_\sigma$.
Hence, by Zorn's Lemma, $\Pp(X)\cap\Ff_\sigma$ has maximal elements.

Now let $S$ be the set of elements of $\Pp(X)\cap\Ff_\sigma$ that
are not below a maximal element.  For any $Y\in S$ there exists a set
$Y'\supsetneq Y$ which must again belong to $S$.  Further, the union
over any chain in $S$ is again contained in $S$.  If $S$ were
non-empty, then, again by Zorn's Lemma, $S$ would contain maximal
elements, which is absurd.\hx

We say that a Muller condition $(\Ff_0,\Ff_1)$ 
is described by a \emph{Zielonka path of co-finite sets} if the
Zielonka tree $Z(\Ff_0,\Ff_1)$ exists, and it is
a finite or infinite path, consisting of co-finite sets,
and possibly the empty set at the end. 

\proposition Every Muller condition on a countable set
$C$, satisfying properties (P0), (P1), and (P2),
is described by a Zielonka path of co-finite sets.
\eprop

\look We already know that the Zielonka tree
for $(\Ff_0,\Ff_1)$ exists.
The set that labels the root of the Zielonka tree is 
$C$ which is co-finite. Consider now any node of the
Zielonka tree, labelled $(X,1-\sigma)$.
If all subsets of $X$ belong to $\Ff_{1-\sigma}$ (in particular
if $X=\emptyset$), then the node is a leaf of the Zielonka tree. 
Otherwise, by Proposition~\ref{prop:max},
we know that $\Pp(X)\cap\Ff_\sigma$ contains a maximal element
$Y$.  
If $X\setminus Y$ was infinite then one could consider any infinite
descending chain $X_1\supsetneq X_2 \supsetneq\dots$ of sets 
in $\Pp(X)\cap\Ff_{1-\sigma}$ whose
intersection is $Y$. But, unless $Y=\emptyset$, this would 
violate property (P1). Hence, $Y$ is either co-finite or empty.
Now suppose that there are two distinct maximal elements $Y_1$, $Y_2$ in
$\Pp(X)\cap\Ff_\sigma$. Since $Y_1,Y_2$ are both co-finite, $Y_1\cap
Y_2\not=\emptyset$. By property (P0), $Y_1\cup Y_2\in \Ff_\sigma$ which is
impossible by the maximality of $Y_1$ and $Y_2$. 
This means that the node $(X,1-\sigma)$ has a unique sucessor $(Y,\sigma)$,
with  $Y$ being the greatest element in $\Pp(X)\cap\Ff_\sigma$.

Thus, the Zielonka tree is indeed a finite or infinite path of
co-finite sets and, if it is finite, with possibly the
empty set at the end.
\hx

Next we have to make precise what it means that a
Muller condition reduces to a parity condition.

\defn A Muller condition $(\Ff_0,\Ff_1)$ 
on $C$, with $\emptyset\in \Ff_\sigma$, reduces
to a parity condition on $\alpha$, if there
is a function $f:C\ra\alpha$ such that,
\begin{enum}
\item for every non-empty $X\subseteq C$ we have that
\[ X\in \Ff_\sigma \ \Iff \ \min f(X) \text{ is even},\]
\item $f^{-1}(d)$ is finite for every $d\in\alpha$,
unless $d=\max f(C)$ and $d$ is even.
\end{enum}
\edefn

Note that such a reduction may have to switch the role
of the two players. Indeed, if $\emptyset\in\Ff_1$,
then the role of Player~1 in the Muller game
must be taken by Player~0 in the parity game since,
by convention, a play of a parity game in which
no priority is seen infinitely often is won by Player~0.

\proposition If a Muller condition $(\Ff_0,\Ff_1)$ reduces to
a parity game on some $\alpha\leq \omega$, then
it guarantees positional winning strategies. \eprop

\look Since $(\Ff_0,\Ff_1)$ guarantees positional
winning strategies if, and only if, $(\Ff_1,\Ff_0)$
does, we may assume that $\emptyset\in \Ff_0$.
With the function $f:C\ra \alpha$ we can
relabel any Muller game $\Gg$ with winning condition
$(\Ff_0,\Ff_1)$ to a parity game $\Gg'$ on the same
game graph. Since $\Gg'$ is positionally determined
it suffices to show that every play $\pi$ in $\Gg$ is
won by the same player as the corresponding
play in $\Gg'$. Let $X$ be the set of priorities
occurring infinitely often in $\pi$.
In the corrsponding play in $\Gg'$,
the set of priorities seen infinitely
often is either $f(X)$ or $f(X)\cup\{d\}$,
where $d=\max f(C)$ is even. 
If $X$ is empty, then so is $f(X)$, and hence
the minimal priority seen infinitely often
in $\pi'$ is either $d$, which is even, or does
not exist. Otherwise, $X\in\Ff_0$ if, and only if,
$\min f(X)$ is even. Hence Player~0 wins $\pi$
if, and only if, she wins $\pi'$.   
\hx
 
We want to prove that any Muller condition 
described by a Zielonka path of co-finite sets
reduces to a parity condition. Before we do
so, we illustrate the reduction by two examples.
First, consider the case that, for certain $a,b,c,d\in\omega$,
\begin{align*}
\Ff_0&=\{ X\subseteq\omega: a\in X\lor b\in X\lor\{c,d\}\cap X=\emptyset\}\\
\Ff_1&=\{X\subseteq\omega: \{a,b\}\cap X=\emptyset\land (c\in X\lor d\in X)\}
\end{align*}
The Zielonka path for $(\Ff_0,\Ff_1)$ is
\[ (\omega,0) \lra (\omega\setminus\{a,b\},1) \lra 
(\omega\setminus\{a,b,c,d\},0)\]
and we can immediately read off an appropriate reduction 
$f$ from $(\Ff_0,\Ff_1)$ 
to a parity condition with three priorities, namely $f(a)=f(b)=0$,
$f(c)=f(d)=1$, and $f(x)=2$ for all other $x\in\omega$.
However, if we change the condition just a little bit, by
moving the empty set from $\Ff_0$ to $\Ff_1$, the
reduction becomes quite different.
The Zielonka path now has the form
\[ (\omega,0) \lra (\omega\setminus\{a,b\},1) \lra 
(\omega\setminus\{a,b,c,d\},0)\lra (\emptyset,1).\]
Since $\emptyset\in\Ff_1$ we have to change the role of
the players. Moreover, we can no longer map
all elements of $\omega\setminus\{a,b,c,d\}$ to the
same priority since a play in the Muller game
may see this set infinitely often without
seeing any of of its elements more than a finite
number of times. Hence an appropriate reduction
$f:\omega\ra\omega$ is now defined by
\[ f(x)=\begin{cases} 1&\text{ for }x=a \text{ and }x=b\\
2&\text{ for }x=c\text{ and }x=d\\
2x+3&\text{ for }x\in\omega\setminus\{a,b,c,d\}.\end{cases}\]

\proposition Every Muller condition that is described by
a Zielonka path of co-finite sets
reduces to a parity game on an ordinal $\alpha\leq\omega$.
\eprop

\look Let $(\Ff_0,\Ff_1)$ be a Muller condition with
$\emptyset\in\Ff_0$. Otherwise we replace 
$(\Ff_0,\Ff_1)$ by $(\Ff_1,\Ff_0)$.

The Zielonka path for $(\Ff_0,\Ff_1)$ gives, for
some $\beta\leq\omega$ a
descending sequence $(Z_i)_{0\leq i< \beta}$ (in case $C\in\Ff_0$)
or $(Z_i)_{1\leq i<\beta}$ (in case $C\in\Ff_1$), which
consists of co-finite sets, and possibly the empty set at the end,
such that 
\begin{itemize}
\item $Z_{2i}\in \Ff_0$, $Z_{2i+1}\in \Ff_1$. 
\item If $Y\subseteq Z_{2i}$
and $Y\not\subseteq Z_{2i+1}$, then $Y\in\Ff_0$.
Similarly, if $Y\subseteq Z_{2i+1}$
and $Y\not\subseteq Z_{2i+2}$, then $Y\in\Ff_1$. 
\end{itemize}

To define the reduction $f: C\ra\alpha$, we distinguish
three cases.
\begin{enum}
\item If the Zielonka path is infinite, 
set $\alpha:=\omega$, and let $f(c)$ be the biggest $i\in\omega$ 
such that $c\in Z_i$. Since $\bigcap_{i\in\omega} Z_i =\emptyset$
this is well-defined.
\item If the Zielonka path is finite and does not
end with the empty set, let $\alpha:=\beta$ and define
$f(c)$ as in the first case.
\item If the Zielonka path is finite, and ends with
  $Z_{2j+2}=\emptyset$, let $\alpha=\omega$ and define $f:C\ra\omega$
  as follows. For $i<2j+1$ we put $f(c)=i$ for all $c\in
  Z_i\setminus Z_{i+1}$.  For $c\in Z_{2j+1}$ we define
  $f(c)$ by means of a bijection from the (infinite) set $Z_{2j+1}$ to
  the set of yet unused odd priorities $\{2n+1: n\geq j\}$.
\end{enum}

Cleary, for any non-empty $X$, we have that 
$X\in \Ff_0$ if, and only if, $\min f(X)$ is even.
Further, $f^{-1}(i)$ is infinite only in the
case that $f^{-1}(i)=Z_i$ is the last set
in the Zielonka path. In that case $i$ is even
and is the maximal element in the range of $f$.
Hence $f$ defines an appropriate reduction
from $(\Ff_0,\Ff_1)$ to a parity condition.
\hx

We can now summarize the characterisation
of the Muller conditions that guarantee positional determinacy.

\theorem\label{thm:muller} For any Muller condition $(\Ff_0,\Ff_1)$ 
over a countable set $C$ of priorities, the following 
are equivalent.
\begin{enum}
\item $(\Ff_0,\Ff_1)$ guarantees positional winning strategies.
\item $\Ff_0$ and $\Ff_1$ are closed under union of chains,
non-empty intersections of chains, and have no strong splits.
\item $(\Ff_0,\Ff_1)$ is described by a Zielonka path of co-finite sets.
\item $(\Ff_0,\Ff_1)$ reduces to a parity
condition on an ordinal $\alpha\leq\omega$.
\end{enum} 
\etheo

%

\usec{Determinacy of Muller Games. }
Theorem~\ref{thm:muller} classifies the Muller conditions
that imply positional determinacy on all game graphs.
We remark that for Muller games, determinacy itself
is an issue that deserves investigation.
If either $\Ff_0$ or $\Ff_1$ is countable, then
the Muller  condition $(\Ff_0,\Ff_1)$ is Borel
(on level ${\boldsymbol \Sigma^0_4}$ or ${\boldsymbol \Pi^0_4}$),
so determinacy follows from Martin's Theorem.
In general however, Muller conditions over countable sets of
priorities need not be Borel.
This can be seen via a simple counting
argument.  There are only $2^{\aleph_0}$ Borel
sets since each of them is described by a countable infinitary formula. 
But there are $2^{2^{\aleph_0}}$ Muller conditions.
Indeed, on the basis
of Boolean Prime Ideal Theorem, which is a weak form
of the Axiom of Choice, it is not too difficult to
construct non-determined Muller games.

\theorem There exist non-determined infinitary Muller games.
\etheo

\look We slightly modify a well-known construction of a non-determined
Gale-Stewart game. The Boolean Prime Ideal Theorem implies that there
exists a free ultrafilter\footnote{An ultrafilter in
  $\langle\Pp(\w),\incl\rangle$ is a set $U\subseteq\Pp(\omega)$ that does
  not contain $\emptyset$, that includes with any set also all its
  supersets, with any two sets also their intersection, and such that
  for any set $x\subseteq\omega$ either $x\in U$ or $\omega\setminus
  x\in U$. An ultrafilter is free if it contains all co-finite
  sets. As a consequence, it does not contain any finite set.}
$U\subseteq\Pp(\omega)$.  Let $\Ff_0=U$ and construct a game graph
such that by playing the game, the players define a strictly
increasing sequence $a_0<a_1<a_2<\dots$, where the numbers $a_{2n}$
are chosen by Player~0, and numbers $a_{2n+1}$ by Player~1, such that
precisely the priorities in $X:=\bigcup_{n\in\omega}
(a_{2n},a_{2n+1}]$ are seen infinitely often.

We claim that the resulting Muller game is not determined.
Assume that Player~0 has a winning strategy $f$ which maps
any increasing sequence $a_0 < a_1 <\dots <a_{2n-1}$ of even
length to $a_{2n}=f(a_0a_1\dots a_{2n-1})>a_{2n-1}$. 
We consider two intertwined counter-strategies 
of Player~1, forcing essentially Player~0 to 
simultaneously perform two plays against himself.
In reply to the first move $a_0$, Player~1 selects
an arbitrary $a_1>a_0$ and then sets up
the two plays as follows: In the first one
she replies to $a_0$ by $a_1$ and  waits for the
answer $a_2=f(a_0a_1)$ by Player~0.
She then uses $a_2$ as her own reply to $a_0$ in the second play
and gets the answer $a_3=f(a_0a_2)$ by Player~0, which she
now uses as her next move in the first play.
There Player~0 responds by $a_4=f(a_0a_1a_2a_3)$ which
is again used by Player~1 as her answer to $a_0a_2a_3$ in
the second play. And so on.

In this way, the two infinite plays result in
sequences $a_0<a_1<a_2<\dots$ and $a_0<a_2<a_3<\dots$.
Since Player~0 plays with his winning strategy in both plays.
it follows that $X=\bigcup_{n\in\omega} (a_{2n},a_{2n+1}] \in U$,
but also $X'=(a_0,a_2]\cup\bigcup_{n>0} (a_{2n+1},a_{2n+2}] \in U$.
By closure under intersection, it follows that $X\cap X' = (a_0,a_1] \in U$.
But $U$ is a free ultrafilter, so it cannot contain a finite set.

By almost precisely the same argument, it also follows that
Player~1 cannot have a winning strategy. 
\hx

\section{Further results}

\subsection{Uncountable sets of priorities}

\noindent In the previous section we have assumed that the set 
of all priorities
is countable.  However, it can be shown that the characterization of the
Muller conditions that guarantee positional winning strategies remains the
same for uncountable sets of priorities.
As in each play, there appear only countably many
priorities, uncountable sets play no role in a Muller condition.
Still, the argument is slightly more involved than in
Theorem~\ref{thm:muller} because we cannot start the construction from
the set $C$ of all priorities. Nevertheless, we can show
that if every restriction 
of the Muller condition to a countable subset of $C$ satisfies (P0), (P1), 
(P2) then it is equivalent to a parity condition.

Let $(\Ff_0,\Ff_1)$ be a Muller condition over an uncountable set 
$C$ of priorities. 

\defn A set $X\in \Ff_0$ is called a \emph{$0$-limit set} if
whenever $X\subseteq X'$ then $X'\in \Ff_0$. 
Similarly for $1$-limit sets.
\edefn

\lemma
There exists a  countable $0$ or $1$-limit set.
\elemma

\look
  If not then we can construct an infinite increasing sequence of
  countable sets such that even indexed sets are from $\Ff_0$ and odd
  indexed sets are from $\Ff_1$.
\hx

\lemma
  If there is a $0$-limit set then there is no $1$-limit set.
\elemma

\look
Otherwise there is a $0$-limit set $X$ and a $1$-limit set $Y$.
Consider $X\cup Y$. By definition of limit sets it
should belong to both $\Ff_0$ and $\Ff_1$.
\hx

\lemma Suppose that $Y\in \Ff_1$ and that $Y\cup \{a\}\in \Ff_0$ is a
$0$-limit set. In this case $\{a\}$ is a $0$-limit set.
\elemma

\look Assume conversely that there is a $Y'\in \Ff_1$ containing $a$.
We have $Y\cup Y'=Y\cup Y'\cup \{a\}\in \Ff_0$ by the assumption
that $Y\cup\{a\}$ is a limit set. Let $Y_1$ be the greatest element
of $\Ff_1$ included in $Y\cup Y'$; it exists by construction from
Thm~\ref{thm:muller}. As $Y,Y'\in\Ff_1$, we have then $Y\subseteq Y_1$ 
and $Y'\subseteq Y_1$, but this is impossible as it implies that
$Y_1=Y\cup Y'\in\Ff_0\cap\Ff_1=\emptyset$.\hx

\lemma For every limit set $X$ there is a priority $a\in X$ such that
$\{a\}$ is a limit set. \elemma

\look Take a limit set $X\in \Ff_0$ and the greatest set $Y\subseteq X$
from $\Ff_1$. The set $D:=X\setminus Y$ is finite by the construction
from Thm~\ref{thm:muller}. Take an arbitrary element $a\in D$. 
If $Y\cup\{a\}$ is a
$0$-limit set then we are done by the previous lemma.  If not then we take
$Y'\in \Ff_1$ containing $Y\cup\{a\}$. Then we take a next priority
$b\in D$ and consider $Y'\cup \{b\}$. As $D$ is finite and $Y\cup
D=X$ is a limit set we can repeat these steps at most $|D|$ number of
times.  \hx

The above lemma allows to introduce the notion of a 
\emph{limit priority}. Let us remove all the limit priorities 
from $C$ and  consider the Muller condition $(\Ff'_0,\Ff'_1)$ obtained by
restricting $(\Ff_0,\Ff_1)$ to this set of priorities. 
Clearly it also satisfies the conditions P1, P2, P3.

\lemma In $(\Ff'_0,\Ff'_1)$ there is no limit set from $\Ff'_0$. \elemma

\look If there were a limit set in $\Ff'_0$ then this set would be
also a limit set with respect to the original condition
$(\Ff_0,\Ff_1)$. But then it would contain a limit priority which is
impossible by the definition of $(\Ff'_0,\Ff'_1)$. \hx

Hence in $(\Ff'_0,\Ff'_1)$ there is a limit set in $\Ff_1$ and we can
choose limit priorities for the other player. Repeat the construction for
$\omega$ steps. We obtain a sequence of nonempty sets of priorities
$A_1,B_1,\dots$.  If after $\omega$ steps the set of remaining priorities is
nonempty then we take a countable subset $R$ of the priorities that are
left and have a decreasing sequence $R_1,R_2,\dots$, where
$R_{2i-1}=\bigcup\set{a_i,b_i,a_{i+1},b_{i+1},\dots}\cup R$ and
$R_{2i}=\bigcup\set{b_i,a_{i+1},b_{i+1},\dots}\cup R$. 
But the existence of such a
sequence contradicts the condition P1.

\subsection{Games of bounded degree}

\noindent The question arises whether the class of winning conditions
that guarantee positional winning strategies becomes larger if we only
consider game graphs of finite degree, or game graphs of finite and bounded
degree.  In particular this question has been asked for
max-parity games and for parity games over larger ordinals than
$\omega$, where the counter-example that we have presented has
infinite degree. It turns out that parity games over $\omega+1$ are
determined, while those over $\omega+2$ are not. However, it seems
quite difficult to give an exact characterisation of the Muller conditions
that guarantee positional winning strategies on all 
game graphs of bounded finite degree.

\proposition Max-parity games with infinitely many priorities
in general do not admit finite memory strategies,
even for solitaire games and even for game graphs
with maximimal degree two.
\eprop

\look Consider the following game where every vertex has degree one or two.

\vspace*{5mm}
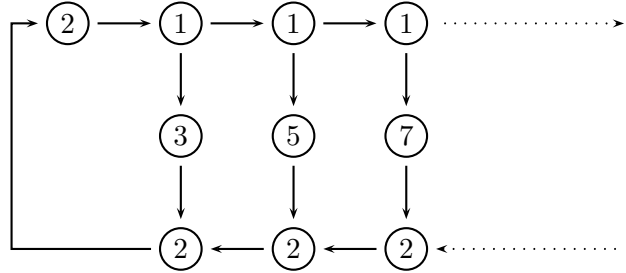
\begin{center}
{\psset{yunit=15mm,xunit=15mm}
\renewcommand{\mcnode}[2]{\circlenode{#1}{#2}}
\begin{pspicture}(5,3)
\rput(0,3){\mcnode{a}{$2$}}
\rput(1,3){\mcnode{b}{$1$}}
\rput(2,3){\mcnode{c}{$1$}}
\rput(3,3){\mcnode{d}{$1$}}
\pnode(5,3){e}
\rput(1,2){\mcnode{bb}{$3$}}
\rput(2,2){\mcnode{cc}{$5$}}
\rput(3,2){\mcnode{dd}{$7$}}
\rput(1,1){\mcnode{bbb}{$2$}}
\rput(2,1){\mcnode{ccc}{$2$}}
\rput(3,1){\mcnode{ddd}{$2$}}
\pnode(5,1){f}
\ncline{a}{b}
\ncline{b}{c}
\ncline{c}{d}
\ncline[linestyle=dotted]{d}{e}
\ncline{b}{bb}
\ncline{c}{cc}
\ncline{d}{dd}
\ncline{bb}{bbb}
\ncline{cc}{ccc}
\ncline{dd}{ddd}
\ncline{ddd}{ccc}
\ncline{ccc}{bbb}
\ncline[linestyle=dotted]{f}{ddd}
\ncbar[angleA=180,angleB=270]{bbb}{a}
\end{pspicture}
}\end{center}
\vspace*{-11 mm}
Assuming the max-parity winning condition it is
obvious that there is an infinite memory strategy
for Player~0 to enforce that
the set of priorities seen infinitely often
is $\{1,2\}$, but that any finite memory
strategy is losing.
\hx

The same construction works for (min-)parity games
on ordinals $\alpha>\omega+1$. Indeed, if we replace
priorities 2 and 1 by $\omega$ and $\omega+1$,
we obtain a min-parity game that requires an infinite
memory winning strategy.

However there is an interesting case where
parity games of bounded degree behave differently
than games of unbounded degree.

\theorem\label{thm:boundeddegree} Parity games of bounded degree
with priorities in $\omega+1$ 
are positionally determined.
\etheo

\look We first consider the case of Player $1$. Extending 
the argument from Theorem~\ref{thm:parity} we show that in any parity
game with priorities in $\omega+1$, Player~1 has a positional winning 
strategy on his winning region. Note that for this case
we do not need the assumption of bounded degree. 

Let $\Gg$ a parity game with priorities in $\omega+1$,
let $g$ be a winning strategy for Player~$1$ on his winning
region in $\Gg$ and let $\Tt_g$ be the associated strategy forest.
In the following, positions labeled by $\w$ will be called \emph{$\w$-positions} and the
other positions will be called \emph{natural positions}.
Since every path through $\Tt_g$ is winning for Player~1,
it must contain infinitely many natural positions. 
This means that the definitions of $1$-signatures that we have
used in the proof for parity games on $\omega$ carry over here. 
Recall that $<^1_i$ denotes the signature order (which is strict and partial). Let
$\tle_i$ be a total well-order extending $<^1_i$, and assume that
$\bot$ is the biggest element in this order.

As in the proof of Theorem~\ref{thm:parity} we associate 
with every node $s$ of $\Tt_g$ of priority $m< \omega$ the $(m+1)$-tuple 
$a(s)=\langle a_0(s),\dots,a_m(s)\rangle\in (\Tt_g\cup\set{\bot})^{m+1}$ of
ancestors. As before, we write $s\prec_m s'$ if there is $i<m$ such
that $a_i(s)\tle_i a_i(s')$ and $a_j(s)= a_j(s')$ for all
$j<i$. Observe that $\prec _m$ is a well-order on vertices of priority
$m$. For a position $v$ of priority $m$ we now take the
$\prec_m$-minimal representant of $v$ in $\Tt_g$, i.e., the minimal $s$
with $h(s)=v$ where $h$ is the canonical homomorphism from $\Tt_g$
to $\Gg$. We denote this representant by $s(v)$.  We can also define the
tuple of ancestors by $a(v)=a(s(v))$.

In order to define $s(v)$ for positions of priority $\omega$ we use
$\prec_m$ to compare vertices in $\Tt_g$ of different, but finite,
priority. We define $s\prec s'$ if either $s \prec_m s'$ for $m$ being
the minimum of the priorities of $s$ and $s'$, or the tuple
$(a_0(s),\dots,a_m(s))$ has $(a_0(s'),\dots,a_{m'}(s'))$ as a strict
prefix. We claim that this is a well-ordering on
vertices of finite priority. To reason by contradiction, suppose that
there is an infinite descending chain $s_0\succ s_1 \succ\dots$ in
this ordering. Let us look at the chain $a_0(s_0),a_0(s_1),\dots$ of
first elements of the tuples. This chain is not increasing in the
$\tle_0$-ordering, so it must eventually stabilise on some element
$a_0$. Let $i_0$ be the position where it stabilises. Observe that
this implies that there cannot be a vertex of priority $0$ after
$s_0$. By a similar argument we find $a_1$ that stabilizes after $a_0$
stabilizes. Continuing like this we get an infinite sequence
$a_0,a_1,\dots$. Observe that infinitely many of the elements in the
sequence are not $\bot$. Indeed, there are vertices of infinitely many
priorities, and when we see a vertex $s$ of priority $i$ then $a_i(s)$
is not $\bot$ so $a_i$ cannot be $\bot$. To finish the argument we
observe that for each $i=0,1,\dots$, if $a_i\not=\bot$ then it is a
vertex of priority $i$ and it is an ancestor of all $a_j\not=\bot$ for
$j>i$. Moreover, there can be no vertices of priority $i$ between
$a_i$ and $a_j\not=\bot$ for $j>i$. Thus, the sequence $a_0,a_1,\dots$
determines an infinite path in $\Tt_g$ where no priority, except
possibly $\omega$, appears infinitely often. This is a contradiction
as we have assumed that all paths in $\Tt_g$ are winning for Player~1.

There are two more notions that we need. For each vertex $s\in\Tt_g$ of
priority $\omega$ we define the \emph{max-distance} to 
be the maximal length of a path of $\omega$-vertices starting from $s$. This
is well defined as on every path from $s$ there
is eventually a vertex of a finite priority. Secondly, for $s$ we 
define its \emph{anchor} to be the closest ancestor of finite
priority. 

Now we are ready to define $s(v)$ for positions $v$ of priority
$\omega$. Among all the representants of $v$, i.e., vertices $s$ such that
$h(s)=v$, we choose one with the $\prec$-smallest anchor. If there are
more than one with this property then among them we choose the one
with the smallest max-distance. If this still does not identify a
unique representant then we choose one arbitrarily.

Having defined $s(v)$ for all $v$ in the winning region for Player $1$
in $\Gg$ we define a positional strategy $g'$. We set $g'(v)=h(t)$
where $t$ is the unique successor of $s(v)$ in $\Tt_g$.  We claim that
this strategy is winning. Suppose conversely that there is a loosing
play respecting the strategy. Then either no natural number appears
infinitely often or the smallest number appearing infinitely often is
even.

If no natural number appears infinitely often then we proceed as in the
proof of Theorem~\ref{thm:infinity}. Consider the suffix of the play after the
last appearance of priority $0$. Let us look at $0$-ancestors of the
positions in this suffix. These ancestors can only get smaller as the
play proceeds. This means that from some moment all positions in
the play will have the same $0$-ancestor. Next, we find a position where 
the $1$-ancestor stabilizes. Observe that it will be a descendant of 
the $0$-ancestor and that there will be no occurrences of priority $0$ between
the two. Proceeding this way we construct a path in the strategy tree
$\Tt_g$ on which no priority, accept possibly $\omega$, appears
on infinite number of times. Notice, that it is important for this
argument that the sequences of $\omega$-nodes are finite.
The remaining case when the smallest number appearing infinitely often
is even is very similar to that from the proof of Theorem~\ref{thm:parity}.

\medskip To show that Player $0$ can win with a positional strategy 
we transform a game $\Gg$ with  priorities from $\omega+1$ into a game $\wh \Gg$ and 
then to $\wt \Gg$, such that $\wt \Gg$ has no $\omega$-positions. Then we
translate the positional winning strategy form $\wt \Gg$ to $\Gg$.

We first describe the transformation from $\Gg$ to $\wh \Gg$. Take a
position $s$ of $\Gg$ labeled with $\omega$. For any $i\in
\omega\cup\set{\omega}$ consider a gadget $K^i_s$: 

 \medskip
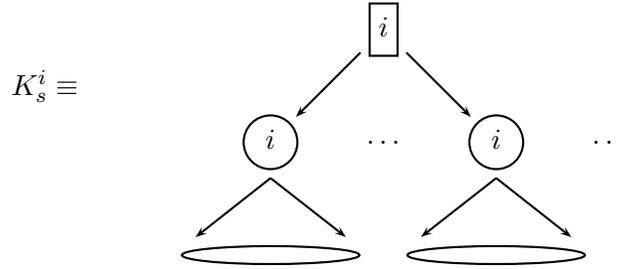
 \begin{center}
 \psset{xunit=1.5cm,yunit=1.5cm,arcangle=15,arrows=->}
 \begin{pspicture}(3,2)
 \rput(2,2){\mbnode{a}{$i$}}
 \rput(1,1){\mcnode{b}{$i$}}
 \rput(3,1){\mcnode{c}{$i$}}
 \rput(1,0){\ovalnode{d}{\hspace{1.5cm}}}
 \rput(3,0){\ovalnode{e}{\hspace{1.5cm}}}
 \rput(2,1){$\ldots$}
 \rput(4,1){$\ldots$}
 \rput(-1,1.5){$K^i_s\equiv$}
 \ncline{a}{b}
 \ncline{a}{c}
 \ncline[offsetB=-1cm]{b}{d}\ncline[offsetB=1cm]{b}{d}
 \ncline[offsetB=-1cm]{c}{e}\ncline[offsetB=1cm]{c}{e}
 \end{pspicture}
 \end{center}

 \medskip

Each round vertex represents a strategy of Player~1 from $s$
permitting him to leave the region of $\w$-labeled positions.  The
oval below such a vertex represents the possible exits, i.e., the
natural positions that Player~0 can reach when Player~1 uses the
chosen strategy. Observe that if Player~0 has a strategy to stay in
$\omega$-positions then the root of $K^\w_s$ has no successors. To be
conform with our definition of the game, in this case we assign a
priority $0$ to the root of $K^\w_s$ and add a self-loop. This way we
make it winning for Player~0. We call such a gadget \emph{degenerate}.

The transformation from $\Gg$ to $\wh \Gg$ is the following. Take an
$\w$-position $s$ and replace it by the gadget $K^\w_s$. 
The leaves of this gadget are natural positions in $\Gg$, 
hence we only add one position of Player~1 and some positions
of Player~0. Redirect every arrow
going from a natural position to $s$ to the root of $K^\w_s$. 
Repeating this for all $\w$-positions (of the game $\Gg$) 
we obtain the game $\wh \Gg$. 
This game has the property that sequences of $\w$-vertices can
have length at most $2$.  Moreover there is an easy correspondence
between strategies in $\Gg$ and $\wh \Gg$.

Next we describe the transformation from $\wh \Gg$ to $\wt \Gg$.  The idea
is to eliminate the priorities $\w$. If we come to a gadget from a
position of priority $i$ then we can as well assume that we see $i$ in
place of $\w$. The result of an infinite play will be the same as we
at most triple the number of $i$'s seen. For example, if $\w$ was the
only priority appearing infinitely often then after the change no
priority at all would appear infinitely often, which gives the win to
the same player.  The transformation from $\wh \Gg$ to $\wt \Gg$ is as
follows.  For each priority $i\in\omega$ and each non-degenerate gadget
$K^\w_s$ we create a gadget $K^i_s$, which has priority $\omega$
replaced by the priority $i$. For each position $u$ of priority $i$ in
$\wh \Gg$ with an arrow to the root of $K^\w_s$ we redirect this arrow
to the root of $K^i_s$. Of course we need not to create $K^i_s$ if
there are no such $u$. Repeating this procedure for each gadget
$K^\w_s$ we get rid of all positions of priority $\w$. The result is
the game $\wt \Gg$.

There is a canonical homomorphism $\wt h : \wt \Gg\to \wh \Gg$ which maps
the root of $K^i_s$ to the root of $K^\w_s$. It should be clear that
there is a winning strategy in $\wt G$ if, and only if, there is one
in $\wh G$ (the image of a path is winning for Player~0 if, and only
if, the path is a winning play for Player~0).  As in $\wt G$ there are
no vertices of priority $\w$ we know that Player~$0$ has a positional
winning strategy on his winning region.  We will show how to translate
it into a positional winning strategy in $G$ (using $\wh G$ on the way).

Take a positional winning strategy $f$ for Player~0 in $\wt G$.
Consider the signature assignment in $\wt G$ defined by the strategy
(as described in the section on parity games). This defines a
signature assignment on natural positions of $\Gg$.  It remains to define
signatures for $\omega$-positions and then use it to define a winning
strategy. For each $\omega$-position $s$ consider the gadget $K^i_s$
for some $i$. If the gadget is degenerated then in $\Gg$ Player~0 has a
strategy from $s$ to stay in $\w$-positions. We are done in this case
as we can assume that Player~0 has one global positional strategy on
all vertices with this property. We call such $s$ \emph{immediately
  winning}.  Suppose then that $K^i_s$ is not degenerate and $f$ is
winning from its root. Each leaf of the gadget has assigned a
signature. We can define the signature of $s$, denoted also by
$\sig^0(s)$, by taking $\inf$ in nodes of Player~0 in $K^i_s$ and then
$\sup$ in $s$. Here $\inf$ and $\sup$ are in the lattice of
$\omega$-vectors of ordinals. Observe that $\sig^0(s)$ does not
depend on the choice of $i$ in $K^i_s$. In order to have a uniform
notation let $\leq^0_\omega$ denote the standard lexicographic
ordering on $\omega$-tuples of ordinals. Notice that this is not a
well-order while $\leq^0_i$ for $i\in\omega$ are.  With this
definition of signatures we have that if $s$ is a position of Player~1,
then for every successor $t$ that is not immediately winning we have
$\sig(t)\leq^0_i \sig(s)$ where $i$ is a priority of $s$. Similarly,
if $s$ is a position of Player~0, then it has a successor $t$ which is either
immediately winning or satisfies the same property.  Having this
property we can define a positional strategy for Player~0 that
consists of choosing the smallest possible signature. The proof that
this strategy is winning is the same as in the case of parity
games. \hx

This theorem indicates that when we limit ourselves to game graphs of
finite degree the class of Muller conditions guaranteeing positional
winning strategies becomes larger. 
There also exist Muller conditions that 
do not reduce to parity conditions over any ordinal but still
guarantee positional winning strategies on all game graphs of 
finite degree. For finite sets of priorities, such examples are
well-known. In the simplest one, the set of priorities 
is $C=\{0,1\}$, with $\Ff_0=\{\{0,1\}\}$ and $\Ff_1=\{\{0\},\{1\}\}$.

Similar examples with an infinite set of colours can be constructed
as follows. Let $Y$ be any infinite set with $e\not\in Y$ and 
set $C=Y\cup \{e\}$. Put
\begin{equation*}
  \Ff_0=\Pp(Y)\cup \set{\set{e}}\cup \set{\es}\qquad \Ff_1=\set{Z :
    e\in Z \land Z\cap    Y\not=\es}
\end{equation*}

It should be clear that $(\Ff_0,\Ff_1)$ is not equivalent to a parity
condition because each priority individually is winning for
Player $0$. By arguments that are similar to
the proof of Theorem~\ref{thm:boundeddegree} one can show 
that such a condition guarantees positional
determinacy on all game graphs of finite degree. 

It is an open problem 
to give a complete characterisation of all such conditions.

\subsection{Finite appearance of priorities}

\noindent We may also ask whether the characterisation of
positionally determined Muller conditions changes
if we only consider games where $\Omega^{-1}(c)$
is finite for every priority $c$.
This is not the case. Indeed, the counter-examples 
for properties (P0) and (P2) are games with this property, and
in the counter-example for (P1) we can easily
eliminate infinite occurrences of priorities.
Consider the figure in the proof of Lemma~\ref{lem:P1}.
It suffices to omit the sets $X_2,X_3,\ldots$
and redirect, for every $i\geq 2$, each arrow from 
$a$ to an element $x_i\in X_i$ to the element $x_i\in X_1$.

\subsection{Related work}

There has recently been some interesting research on similar questions
for games in somewhat different settings.  For instance, Colcombet and
Niwi\'nski \cite{ColcombetNiw06} have studied positional determinacy
of games where edges, rather than vertices are labeled by priorities.
This changes the situation completely. For instance, it is easily seen
that there are edge-labeled parity games with infinitely many
priorities that require winning strategies with infinite memory. Also
there are some very simple non-Muller winning conditions that
guarantee positional determinacy on vertex-labeled games but fail to
do so on edge-labeled ones. An example is the set $(0+1)^*(01)^\omega$
whare Player~0 has to make sure that from some point onwards the
priorities 0 and 1 alternate. If she can achieve this on a
vertex-labelled game then she can also do this positionally. However,
when the priorities are on the edges, then this is not the case:
consider the game with a single vertex and two self-loops with
priorities 0 and 1.  In fact, Colcombet and Niwi\'nski prove that the
only prefix-independent winning conditions that guarantee positional
determinacy on all edge-labeled game graphs are precisely the parity
conditions with a finite number of priorities. In a similar vein,
Kopczynski \cite{Kopczynski06} characterises the winning
conditions that guarantee positional
determinacy \emph{for one player} on edge-labeled game graphs.

Serre~\cite{Serre04} exhibits examples of winning conditions 
on a countable set of priorities that have high Borel complexity, 
but still admit positional winning strategies. Recall that in our setting, 
if the set of priorities is countable then the conditions are at 
most at levels $\Sigma^0_4$ or $\Pi^0_4$ of the Borel hierarchy.

Gimbert and Zielonka \cite{GimbertZiel05} consider edge-labeled games
with real valued pay-offs. They characterise those pay-off functions
that guarantee optimal positional strategies for both players on all
finite game graphs. As in the case studied by Colcombet and Niwi\'nski the
payoffs are on edges and not on vertices.


\begin{thebibliography}{10}

\bibitem{ArnoldVinWal03}
{\sc A.~Arnold, A.~Vincent, and I.~Walukiewicz}, {\em Games for synthesis of
  controllers with partial observation}, Theoretical Computer Science, 303
  (2003), pp.~7--34.

\bibitem{BerwangerDawHunKre06}
{\sc D.~Berwanger, A.~Dawar, P.~Hunter, and S.~Kreutzer}, {\em Dag-width and
  parity games}, in Proceedings of 23rd Annual Symposium on Theoretical Aspects
  of Computer Science, STACS 2006, Lecture Notes in Computer Science Nr.~3848,
  2006, pp.~524--536.

\bibitem{BerwangerGra04}
{\sc D.~Berwanger and E.~Gr{\"a}del}, {\em Fixed-point logics and solitaire
  games}, Theory of Computing Systems, 37 (2004), pp.~675--694.

\bibitem{BerwangerGra05}
{\sc D.~Berwanger and E.~Gr{\"a}del}, {\em Entanglement - {A} measure for the
  complexity of directed graphs with applications to logic and games}, in
  Proceedings of LPAR 2004, Lecture Notes in Computer Science Nr.~3452,
  Springer-Verlag, 2005, pp.~209--223.

\bibitem{BlumensathGra04}
{\sc A.~Blumensath and E.~Gr{\"a}del}, {\em Finite presentations of infinite
  structures: Automata and interpretations}, Theory of Computing Systems, 37
  (2004), pp.~641 -- 674.

\bibitem{BouquetSerWal03}
{\sc A.~Bouquet, O.~Serre, and I.~Walukiewicz}, {\em Pushdown games with
  unboundedness and regular conditions}, in Proceedings of FSTTCS'03, Lecture
  Notes in Computer Science Nr.~2914, 2003, pp.~88--99.

\bibitem{CachatDupTho02}
{\sc T.~Cachat, J.~Duparc, and W.~Thomas}, {\em Solving pushdown games with a
  ${\Sigma_3}$ winning cndition}, in Computer Science Logic, CSL 2002, Lecture
  Notes in Computer Science Nr.~2471, Springer-Verlag, 2002, pp.~322--336.

\bibitem{ColcombetNiw06}
{\sc T.~Colcombet and D.~Niwi\'nski}, {\em On the positional determinacy of
  edge-labeled games}, Theoretical Computer Science,  (2006).

\bibitem{DziembowskiJurWal97}
{\sc S.~Dziembowski, M.~Jurdzi{\'n}ski, and I.~Walukiewicz}, {\em How much
  memory is needed to win infinite games?}, in Proceedings of 12th Annual
  {IEEE} Symposium on Logic in Computer Science (LICS 97), 1997, pp.~99--110.

\bibitem{EmersonJut91}
{\sc A.~Emerson and C.~Jutla}, {\em Tree automata, mu-calculus and
  determinacy}, in Proc. 32nd IEEE Symp. on Foundations of Computer Science,
  1991, pp.~368--377.

\bibitem{EmersonJutSis01}
{\sc A.~Emerson, C.~Jutla, and P.~Sistla}, {\em On model checking for the
  $\mu$-calculus and its fragments}, Theoretical Computer Science, 258 (2001),
  pp.~491--522.

\bibitem{Gimbert04}
{\sc H.~Gimbert}, {\em Parity and exploration games on infinite graphs}, in
  Proceedings of CSL 2004, Lecture Notes in Computer Science Nr.~3210,
  Springer, 2004, pp.~56--70.

\bibitem{GimbertZiel05}
{\sc H.~Gimbert and W.~Zielonka}, {\em Games where you can play optimally
  without any memory}, in CONCUR 2005 - Concurrency Theory, 16th International
  Conference, Lecture Notes in Computer Science Nr.~3653, Springer, 2005,
  pp.~428--442.

\bibitem{Graedel06}
{\sc E.~Gr{\"a}del}, {\em Finite {M}odel {T}heory and {D}escriptive
  {C}omplexity}, in Finite Model Theory and Its Applications, Springer-Verlag,
  2006.
\newblock To appear.

\bibitem{GraedelThoWil02}
{\sc E.~Gr{\"a}del, W.~Thomas, and T.~Wilke}, eds., {\em Automata, Logics, and
  Infinite Games}, Lecture Notes in Computer Science Nr.~2500, Springer, 2002.

\bibitem{GurevichHar82}
{\sc Y.~Gurevich and L.~Harrington}, {\em Trees, automata and games}, in
  Proceedings of the 14th Annual ACM Symposium on Theory of Computing,
  STOC~'82, 1982, pp.~60--65.

\bibitem{Jurdzinski98}
{\sc M.~Jurdzi\'nski}, {\em Deciding the winner in parity games is in {UP}
  $\cap$ {C}o-{UP}}, Information Processing Letters, 68 (1998), pp.~119--124.

\bibitem{Jurdzinski00}
{\sc M.~Jurdzi{\'n}ski}, {\em Small progress measures for solving parity
  games}, in Proceedings of 17th Annual Symposium on Theoretical Aspects of
  Computer Science, STACS 2000, Lecture Notes in Computer Science Nr.~1770,
  Springer, 2000, pp.~290--301.

\bibitem{JurdzinskiPatZwi06}
{\sc M.~Jurdzi{\'n}ski, M.~Paterson, and U.~Zwick}, {\em A deterministic
  subexponential algorithm for solving parity games}, in Proceedings of
  ACM-SIAM Proceedings on Discrete Algorithms, SODA 2006, 2006, pp.~117--123.

\bibitem{Kopczynski06}
{\sc E.~Kopczynski}, {\em Half-positional determinacy of infinite games}, in
  Automata, Languages and Programming, 33rd International Colloquium, ICALP
  2006, Lecture Notes in Computer Science Nr.~4052, 2006, pp.~336--347.

\bibitem{KupfermanVar00}
{\sc O.~Kupferman and M.~Vardi}, {\em An automata-theoretic approach to
  reasoning about infinite-state systems}, in Proceedings of 12th International
  Conference on Computer-Aided Verification CAV 2000, Lecture Notes in Computer
  Science Nr.~1855, Springer, 2000, pp.~36--52.

\bibitem{Martin75}
{\sc D.~Martin}, {\em Borel determinacy}, Annals of Mathematics, 102 (1975),
  pp.~336--371.

\bibitem{McNaughton93}
{\sc R.~McNaughton}, {\em Infinite games played on finite graphs}, Annals of
  Pure and Applied Logic, 65 (1993), pp.~149--184.

\bibitem{Mostowski91}
{\sc A.~Mostowski}, {\em Games with forbidden positions}, Tech. Rep. Tech.
  Report 78, University of Gdansk, 1991.

\bibitem{Obdrzalek03}
{\sc J.~Obdrzalek}, {\em Fast mu-calculus model checking when tree-width is
  bounded}, in Proceedings of CAV 2003, vol.~2752 of LNCS, Springer, 2003,
  pp.~80--92.

\bibitem{Obdrzalek06}
{\sc J.~Obdrzalek}, {\em {DAG}-width - connectivity measure for directed
  graphs}, in Proceedings of ACM-SIAM Proceedings on Discrete Algorithms, SODA
  2006, 2006, pp.~814--821.

\bibitem{Serre04}
{\sc O.~Serre}, {\em Games with winning conditions of high {B}orel complexity},
  in Proceedings of ICALP 2004, vol.~3142 of Lecture Notes in Computer Science,
  2004, pp.~1150--1162.

\bibitem{Walukiewicz01}
{\sc I.~Walukiewicz}, {\em Pushdown processes: {G}ames and model checking},
  Information and Computation, 164 (2001), pp.~234--263.

\bibitem{Zielonka98}
{\sc W.~Zielonka}, {\em Infinite games on finitely coloured graphs with
  applications to automata on infinite trees}, Theoretical Computer Science,
  200 (1998), pp.~135--183.

\end{thebibliography}

\end{document}